\documentclass[twoside,12pt]{article}
\usepackage[T1]{fontenc}
\usepackage[latin9]{inputenc}
\usepackage{amsmath}
\usepackage{amssymb}
\usepackage{graphicx}
\usepackage{CJK}
\usepackage{amsmath}
\usepackage{cases}
\usepackage{indentfirst}
\usepackage{cite}
\usepackage{bm}
\usepackage{epsfig}

\newenvironment{sequation}{\begin{equation}\small}{\end{equation}}

\begin{document}
\title{\textbf{The Directed Dominating Set problem studied by cavity method:Warning propagation and population dynamics}}
\author{$Yusupjan Habibulla$\\ School of Physics and Technology, xinjiang university, Sheng-Li Road 14,\\  Urumqi 830046, China}

\maketitle
\tableofcontents
\begin{abstract}
The minimal dominating set for a digraph(directed graph)is a prototypical hard combinatorial optimization problem. In a previous paper, we studied this problem using the cavity method. Although we found a solution for a given graph that gives very good estimate of the minimal dominating size, we further developed the one step replica
symmetry breaking theory to determine the ground state energy of the undirected minimal dominating set problem. The solution space for the undirected minimal dominating set problem exhibits both condensation transition and cluster transition on regular random graphs. We also developed the zero temperature survey propagation algorithm on undirected Erd\H{o}s-R$\acute{e}$nyi graphs to find the ground state energy. In this paper we continue to develope the one step replica
symmetry breaking theory to find the ground state energy for the directed minimal dominating set problem. We find the following. (1)The warning propagation equation can not converge when the connectivity is greater than the core percolation threshold value of 3.704. Positive edges have two types warning, but the negative edges have one. (2)We determine the ground state energy and the transition point of the Erd\H{o}s-R$\acute{e}$nyi random graph. (3)The survey propagation decimation algorithm has good results comparable with the belief propagation decimation algorithm.\\\\
\textbf{\large Keywords: }directed minimal dominating set , replica
symmetry breaking, Erd\H{o}s-R$\acute{e}$nyi graph, warning propagation, survey propagation decimation.
\end{abstract}
\section{ Introduction}
The minimum dominating set for a general digraph\cite{1,2} is a fundamental nondeterministic polynomial-hard (NPhard)
combinatorial optimization problem \cite{3}. Any digraph $D = {V,A}$ contains a set $V\equiv{1, 2, \ldots,N}$ of $N$ vertices and a set $A\equiv\{(i, j) : i, j\in V \}$ of $M$ arcs(directed edges), where each arc $(i, j)$ points from a parent vertex (predecessor) $i$ to a child vertex (successor) $j$. The arc density $\alpha$ is defined by $\alpha\equiv M/N$. A directed dominating set $\Gamma$ is a vertex set such that, for any node in the network, either the node itself or  one of its predecessors belongs to 
$\Gamma$.
a directed minimal dominating set(DMDS) is a smallest directed dominating set. The DMDS problem is very important for monitoring and controlling the directed interaction processes\cite{4,5,6,7,8,9,10} in the complex networks.\\
The Statistical physicists have widely studied this optimization problem using the statistical physics of spin glass systems\cite{11,12,13,14,15,16,17,18,19}. The cavity method is used to estimate the occupation probability at each node, and a solution for the given problem is constructed using this probabilities. Using statistical physicis does not always yield a single solution for a given graph, but if this graph does not contain any small loops, then the cavity equation may converge to a stable point, so we can study the problem using the cavity method. \\
Research into spin glass systems can be divided into two levels-the replica symmetry (RS) level and the replica symmetry breaking (RSB) level. For a given optimization problem, RS theory finds the smallest set in the given graph that satisfy this problem, and RSB theory finds the number of solutions and the ground state energy. Previously, we have found the smallest minimal dominating set (MDS) using the cavity method on both directed and undirected networks\cite{20,21}, and have found the ground state energy for undirected networks\cite{22}. The current study is inspired by \cite{22}, where we studied the ground state energy for the DMDS problem also using the cavity method. At a temperature of zero, we used survey propagation to find the ground state energy of the Erd\H{o}s-R$\acute{e}$nyi graph (ER) random graph.\\
This paper studies the solution space of the DMDS problem using one step replica symmetry breaking(1RSB) theory from statistical physics. This work is a continuation of our earlier work\cite{22} on the solution space of the undirected MDS problem.
We organize the paper as follows. In section 2, we recall RS theory for spin glass sytems before introducing RSB theory. We present the belief propagation (BP) equation, thermodynamic quantities and warning propagation for the DMDS problem. In the section 3, we introduce1RSB theory and the associated thermodynamic quantities. We then derive 1RSB theory and thermodynamic quantities at $\beta=\infty$, and survey propagation(SP) for the DMDS problem, before introducing the survey propagation decimation (SPD) process for the DMDS problem in detail. Finally in Section 4 we conclusion our results.

\section{Replica Symmetry}
\subsection{General Replica Symmetry Theory}
To estimate the MDS for a given graph using the way of mean field theory, we must have the partition function for the given problem. The partition function Z is
\begin{equation}
Z=\sum_{\underline{c}}\prod_{i\in W}{e^{-\beta c_{i}}}[1-(1-c_{i})\prod_{k\in\partial i^{+}}(1- c_{k})]
\end{equation}
where $\underline{ c}\equiv( c_{1}, c_{2},\ldots, c_{n})$denotes one of the $2^{n}$ possible configurations, $ c_{i}=+1$if node $i$ is occupied and $ c_{i}=0$ otherwise,$\beta$ is the inverse temperature, and $\partial i$ denotes the neighbors of node $i$. The partition function therefore only takes into account the dominating sets.\\
We use RS mean field theory, such as the Bethe-Peierls approximation\cite{23} or partition function expansion\cite{24,25}, to solve the above spin glass model. We assign the cavity message $p_{i\rightarrow j}^{(c_{i},c_{j})}$ on the every edge,and these messages must satisfy the equation

\begin{equation}
p_{i\rightarrow j}^{(c_{i},c_{j})}=\frac{e^{-\beta c_{i}}[\prod\limits_{k\in\partial i^{+}}\sum\limits_{c_{k}}p_{k\rightarrow i}^{c_{k}, c_{i}}-\delta_{ c_{i}}^{0}\prod\limits_{k\in\partial i^{+}}p_{k\rightarrow i}^{0,0}]\prod\limits_{l\in\partial i^{-}\backslash j}\sum\limits_{c_{l}}p_{l\leftarrow i}^{c_{l}, c_{i}}}{\sum\limits_{\acute{ c}_{i},\acute{ c}_{j}}e^{-\beta \acute c_{i}}[\prod\limits_{k\in\partial i^{+}}\sum\limits_{\acute c_{k}}p_{k\rightarrow i}^{\acute c_{k}, \acute c_{i}}-\delta_{ \acute{ c}_{i}}^{0}\prod\limits_{k\in\partial i^{+}}p_{k\rightarrow i}^{0,0}]\prod\limits_{l\in\partial i^{-}\backslash j}\sum\limits_{c_{l}}p_{l\leftarrow i}^{c_{l}, c_{i}}}
\end{equation}
\begin{equation}
p_{i\leftarrow j}^{(c_{i},c_{j})}=\frac{e^{-\beta c_{i}}[\prod\limits_{k\in\partial i^{+}\backslash j}\sum\limits_{c_{k}}p_{k\rightarrow i}^{c_{k}, c_{i}}-\delta_{ c_{i}+c_{j}}^{0}\prod\limits_{k\in\partial i^{+}\backslash j}p_{k\rightarrow i}^{0,0}]\prod\limits_{l\in\partial i^{-}}\sum\limits_{c_{l}}p_{l\leftarrow i}^{c_{l}, c_{i}}}{\sum\limits_{\acute{ c}_{i},\acute{ c}_{j}}e^{-\beta \acute c_{i}}[\prod\limits_{k\in\partial i^{+}\backslash j}\sum\limits_{\acute c_{k}}p_{k\rightarrow i}^{\acute c_{k}, \acute c_{i}}-\delta_{ \acute{ c}_{i}+\acute{ c}_{j}}^{0}\prod\limits_{k\in\partial i^{+}\backslash j}p_{k\rightarrow i}^{0,0}]\prod\limits_{l\in\partial i^{-}}\sum\limits_{c_{l}}p_{l\leftarrow i}^{c_{l}, c_{i}}}
\end{equation}
known as the BP equation. The Kronecker symbol is defined by $\delta_{m}^{n}=1$ if $m=n$ and $\delta_{m}^{n}=0$ otherwise. The cavity message $p_{i\rightarrow j}^{(c_{i},c_{j})}$ represents the joint probability that node $i$ is in state $c_{i}$ and the adjacent node $j$ is in state $c_{j}$ when the constraint for node $j$ is not considered.The marginal probability $p_{i}^{c}$ for node $i$ is expressed as
\begin{equation}
p_{i}^{c}=\frac{e^{-\beta c}[\prod\limits_{j\in\partial i^{+}}\sum\limits_{c_{j}}p_{j\rightarrow i}^{c_{j}, c}-\delta_{ c}^{0}\prod\limits_{j\in\partial i^{+}}p_{j\rightarrow i}^{0,0}]\prod\limits_{k\in\partial i^{-}}\sum\limits_{c_{k}}p_{k\leftarrow i}^{c_{k}, c}}{\sum\limits_{c_{i}}e^{-\beta c_{i}}[\prod\limits_{j\in\partial i^{+}}\sum\limits_{c_{j}}p_{j\rightarrow i}^{c_{j}, c_{i}}-\delta_{ c_{i}}^{0}\prod\limits_{j\in\partial i^{+}}p_{j\rightarrow i}^{0,0}]\prod\limits_{k\in\partial i^{-}}\sum\limits_{c_{k}}p_{k\leftarrow i}^{c_{k}, c_{i}}}
\end{equation}

Finally the free energy can be calculated using mean field theory as
\begin{equation}
F_{0}=\sum_{i=1}^{N}F_{i}-\sum_{(i,j)=1}^{M}F_{(i,j)}
\end{equation}

where

\begin{equation}
F_{i}=-\frac{1}{\beta}\ln\{\sum\limits_{c_{i}}e^{-\beta c_{i}}[\prod\limits_{j\in\partial i^{+}}\sum\limits_{c_{j}}p_{j\rightarrow i}^{c_{j}, c_{i}}-\delta_{ c_{i}}^{0}\prod\limits_{j\in\partial i^{+}}p_{j\rightarrow i}^{0,0}]\prod\limits_{k\in\partial i^{-}}\sum\limits_{c_{k}}p_{k\leftarrow i}^{c_{k}, c_{i}}\}
\end{equation}

\begin{equation}
F_{(i,j)}=-\frac{1}{\beta}\ln[\sum_{ c_{i}, c_{j}}p_{i\rightarrow j}^{c_{i}, c_{j}}p_{j\leftarrow i}^{c_{j}, c_{i}}]
\end{equation}
We use $F_{i}$ to denote the free energy at node $i$, and $F_{(i,j)}$ to denote the free energy of the edge $(i,j)$.We iterate the BP equation until it converges to a stable point, and then calculate the mean free energy $f\equiv F/N$ and the energy density $\omega=1/N\sum_{i}p_{i}^{+1}$ using equation (3) and (4). The entropy density is calculated as $s=\beta(\omega-f)$.
\subsection{Warning Propagation }
In this section, we introduce BP at $\beta=\infty$, which is called warning propagation. Even though the warning propagation may converge very quickly, it can only converge for $C<3.704$ on an ER random network, so we must further consider the 1RSB case at $\beta=\infty$.
To estimate the minimal energy of an MDS, we must consider the limit at $\beta=\infty$. There are three cases for a single node: (1) node $i$ appears in every MDS, namely $p_{i}^{1}=1, p_{i}^{0}=0$; (2) node $i$ appears in no MDS, namely $p_{i}^{1}=0, p_{i}^{0}=1$; (3) node $i$ appears in some but not all MDSs, namely $p_{i}^{1}=0.5, p_{i}^{0}=0.5$. Thus there are nine cases for the pair of nodes $(i,j)$, However, only four cavity messages are possible: (1) node $i$ appears in every MDS and node $j$ appears in some but not all MDSs, namely $p_{i\rightarrow j}^{1,0}=p_{i\rightarrow j}^{1,1}=0.5, p_{i\rightarrow j}^{0,0}=p_{i\rightarrow j}^{0,1}=0$; (2) node $i$ appears in no MDS and node $j$ appears in some but not all MDSs, namely $p_{i\rightarrow j}^{1,0}=p_{i\rightarrow j}^{1,1}=0, p_{i\rightarrow j}^{0,0}=p_{i\rightarrow j}^{0,1}=0.5$; (3) node $i$ appears in no MDS and node $j$ appears in every MDS, namely $p_{i\rightarrow j}^{1,0}=p_{(i,j)}^{1,1}=0, p_{i\rightarrow j}^{0,0}=0,p_{i\rightarrow j}^{0,1}=1$; (4) node $i$ appear in some but not all MDSs and node $j$ appears in some but not all MDSs, namely $p_{i\rightarrow j}^{1,0}=p_{i\rightarrow j}^{1,1}=p_{(i,j)}^{0,0}=p_{i\rightarrow j}^{0,1}=0.25$.\\
The other five cases do not satisfy the normalization condition: (1) if node $i$ appears in every MDS and node $j$ appears in every MDS, namely $p_{i\rightarrow j}^{1,1}=1.0$, then from the relationship $p_{i\rightarrow j}^{1,0}=p_{i\rightarrow j}^{1,1}$ we can derive $p_{i\rightarrow j}^{1,0}=1.0$ so that the total probability is greater than $1$, which is not possible; (2) if node $i$ appears in every MDS and node $j$ appears in no MDS, namely $p_{i\rightarrow j}^{1,0}=1.0$, then in the same way we can derive $p_{i\rightarrow j}^{1,1}=1.0$, so the total probability is again greater than $1$; (3) if node $i$ appears in some but not all MDSs and node $j$ appears in every MDS, namely $p_{i\rightarrow j}^{0,1}=p_{i\rightarrow j}^{1,1}=0.5$, then $p_{i\rightarrow j}^{1,0}=0.5$; (4) if node $i$ appears in some but not all MDSs and node $j$ appears in no MDS, namely $p_{i\rightarrow j}^{0,0}=p_{i\rightarrow j}^{1,0}=0.5$, then $p_{i\rightarrow j}^{1,1}=0.5$; (5) if node $i$ appears in no MDS and node $j$ appears in no MDS, namely $p_{i\rightarrow j}^{0,0}=1$, then from equations (2) and (3) we see that $p_{i\rightarrow j}^{0,0}$ is always smaller than $p_{i\rightarrow j}^{0,1}$ and does not exceed 0.5. It is good to understand that if a node is not occupied, then it cannot request any of its neighbors to be unoccupied in the MDS problem. On the other hand, if a node is occupied, then it cannot request any of its neighbors to be either occupied or unoccupied. 
There is one warning message $p_{i\rightarrow j}=0$ for a single node, but there are two warning messages for a pair nodes $(i,j)$, that is, $p_{i\rightarrow j}^{0,1}=1$ and $p_{i\rightarrow j}^{0,1}=0.5$. The warning message $p_{i\rightarrow j}^{0,1}=1$ is called a first type warning, and the warning message $p_{i\rightarrow j}^{0,1}=0.5$ is called a second type warning.
\begin{equation}p_{i\leftarrow j}^{0,1}=
\begin{cases}
0.0&\qquad\sum\limits_{k\in\partial i^{+}\backslash j}\delta_{p_{k\rightarrow i}^{0,1}}^{1}+\sum\limits_{k\in\partial i^{-}}\delta_{p_{k\leftarrow i}^{0,1}}^{1}\geq2\\
0.25&\qquad\sum\limits_{k\in\partial i^{+}\backslash j}\delta_{p_{k\rightarrow i}^{0,1}}^{1}+\sum\limits_{k\in\partial i^{-}}\delta_{p_{k\leftarrow i}^{0,1}}^{1}=1\\
0.5&\qquad\sum\limits_{k\in\partial i^{+}\backslash j}\delta_{p_{k\rightarrow i}^{0,1}}^{1}+\sum\limits_{k\in\partial i^{-}}\delta_{p_{k\leftarrow i}^{0,1}}^{1}=0\qquad and \qquad\sum\limits_{k\in\partial i^{+}\backslash j}\delta_{p_{k\rightarrow i}^{0,1}}^{0.5}<k^{+}-1\\
1.0&\qquad\sum\limits_{k\in\partial i^{+}\backslash j}\delta_{p_{k\rightarrow i}^{0,1}}^{1}+\sum\limits_{k\in\partial i^{-}}\delta_{p_{k\leftarrow i}^{0,1}}^{1}=0\qquad and \qquad\sum\limits_{k\in\partial i^{+}\backslash j}\delta_{p_{k\rightarrow i}^{0,1}}^{0.5}=k^{+}-1
\end{cases}
\end{equation}

If node $i$ is not covered or observed, so the neighbor node $j$ must be covered, the corresponding case is $p_{i\leftarrow j}^{0,1}=1.0$. If node $i$ is not covered, but it has been observed, so node $j$ can be covered or uncovered, namely $p_{i\leftarrow j}^{0,1}=0.5$. So we can easily understand the upper equation. In the first line, if two or more neighbors(successors exactly) of the node $i$ are not covered or observed, then we must cover the node $i$ to observe them, namely $p_{i\leftarrow j}^{0,1}=p_{i\leftarrow j}^{0,0}=0.0$ or $p_{i}^{0}=0$. In the second line, if only one neighbor(successor exactly) of the node $i$ is not covered or observed, then we have both opportunity to cover or uncover the node $i$, namely $p_{i\leftarrow j}^{0,1}=p_{i\leftarrow j}^{0,0}=0.25$ or $p_{i}^{0}=0.5$. In the third line, if every successor of the node $i$ is observed , but it's observed predecessors less than $k^{+}-1$, then we must not cover the node $i$, the neighbor $j$ can be covered or uncovered, namely $p_{i\leftarrow j}^{0,1}=p_{i\leftarrow j}^{0,0}=0.5$ or $p_{i}^{0}=1.0$. In the fourth line, if every successor of the node $i$ is observed, but the every predecessors of the node $i$ are observed except node $j$, so we must not cover the node $i$, and the neighbor $j$ must be covered, namely $p_{i\leftarrow j}^{0,1}=1.0$ or $p_{i}^{0}=1.0$. We can find that only the successor nodes provide the first type warning, in the same way we can read the following equation

\begin{equation}p_{i\rightarrow j}^{0,1}=
\begin{cases}
0.0&\qquad\sum\limits_{k\in\partial i^{+}}\delta_{p_{k\rightarrow i}^{0,1}}^{1}+\sum\limits_{k\in\partial i^{-}\backslash j}\delta_{p_{k\leftarrow i}^{0,1}}^{1}\geq2\\
0.25&\qquad\sum\limits_{k\in\partial i^{+}}\delta_{p_{k\rightarrow i}^{0,1}}^{1}+\sum\limits_{k\in\partial i^{-}\backslash j}\delta_{p_{k\leftarrow i}^{0,1}}^{1}=1\\
0.5&\qquad\sum\limits_{k\in\partial i^{+}}\delta_{p_{k\rightarrow i}^{0,1}}^{1}+\sum\limits_{k\in\partial i^{-}\backslash j}\delta_{p_{k\leftarrow i}^{0,1}}^{1}=0\qquad and \qquad\sum\limits_{k\in\partial i^{+}}\delta_{p_{k\rightarrow i}^{0,1}}^{0.5}<k^{+}\\
0.25&\qquad\sum\limits_{k\in\partial i^{+}}\delta_{p_{k\rightarrow i}^{0,1}}^{1}+\sum\limits_{k\in\partial i^{-}\backslash j}\delta_{p_{k\leftarrow i}^{0,1}}^{1}=0\qquad and \qquad\sum\limits_{k\in\partial i^{+}}\delta_{p_{k\rightarrow i}^{0,1}}^{0.5}=k^{+}
\end{cases}
\end{equation}

The above equations are called warning propagation equations. Only the message $p_{i\leftarrow j}^{0,1}$ can produce a first type warning when the incoming messages do not include any first type messages and the incoming positive messages have $\partial i^{+}-1$ second type messages. If we find the stable point of the warning propagation,then we can calculate the coarse-grained state of every node as
\begin{equation}p_{i}^{1}=
\begin{cases}
0.0&\qquad\sum\limits_{k\in\partial i^{+}}\delta_{p_{k\rightarrow i}^{0,1}}^{1}+\sum\limits_{k\in\partial i^{-}}\delta_{p_{k\leftarrow i}^{0,1}}^{1}=0\qquad and \qquad\sum\limits_{k\in\partial i^{+}}\delta_{p_{k\rightarrow i}^{0,1}}^{0.5}<k^{+}\\
0.5&\qquad\sum\limits_{k\in\partial i^{+}}\delta_{p_{k\rightarrow i}^{0,1}}^{1}+\sum\limits_{k\in\partial i^{-}}\delta_{p_{k\leftarrow i}^{0,1}}^{1}=0\qquad and \qquad\sum\limits_{k\in\partial i^{+}}\delta_{p_{k\rightarrow i}^{0,1}}^{0.5}=k^{+}\\
0.5&\qquad\sum\limits_{k\in\partial i^{+}}\delta_{p_{k\rightarrow i}^{0,1}}^{1}+\sum\limits_{k\in\partial i^{-}}\delta_{p_{k\leftarrow i}^{0,1}}^{1}=1\\
1.0&\qquad\sum\limits_{k\in\partial i^{+}}\delta_{p_{k\rightarrow i}^{0,1}}^{1}+\sum\limits_{k\in\partial i^{-}}\delta_{p_{k\leftarrow i}^{0,1}}^{1}\geq2
\end{cases}
\end{equation}
and we can calculate the free energy for the DMDS problem in the general case as
\begin{equation}
\begin{split}
F_{i}&=-\frac{1}{\beta}\ln\{[\prod_{k\in\partial i^{+}}(p_{k\rightarrow i}^{0,0}+p_{k\rightarrow i}^{1,0})-\prod_{k\in\partial i^{+}}p_{k\rightarrow i}^{0,0}]*\prod_{k\in\partial i^{-}}(p_{k\leftarrow i}^{0,0}+p_{k\leftarrow i}^{1,0})\\
&+e^{-\beta}\prod_{k\in\partial i^{+}}(p_{k\rightarrow i}^{0,1}+p_{k\rightarrow i}^{1,1})\prod_{k\in\partial i^{-}}(p_{k\leftarrow i}^{0,1}+p_{k\leftarrow i}^{1,1})\}
\end{split}
\end{equation}

\begin{equation}
\begin{split}
F_{ij}&=-\frac{1}{\beta}\ln(p_{i\rightarrow j}^{0,0}p_{j\leftarrow i}^{0,0}+p_{i\rightarrow j}^{0,1}p_{j\leftarrow i}^{1,0}\\
&\quad+p_{i\rightarrow j}^{1,0}p_{j\leftarrow i}^{0,1}+p_{i\rightarrow j}^{1,1}p_{j\leftarrow i}^{1,1})
\end{split}
\end{equation}
The energy equals the free energy when $\beta=\infty$, so from the above questions we can write the free energy and the energy as 
\begin{equation}
\begin{split}
E_{min}&=\lim\limits_{\beta\rightarrow\infty}F_{0}=\sum\limits_{i=1}^{N}[\Theta[(\sum\limits_{j\in\partial i^{+}}\delta_{p_{j\rightarrow i}^{0,1}}^{1}+\sum\limits_{j\in\partial i^{-}}\delta_{p_{j\leftarrow i}^{0,1}}^{1})-1]+\delta(\sum\limits_{j\in\partial i^{+}}\delta_{p_{j\rightarrow i}^{0,1}}^{0.5},k^{+})]\\
&-\sum\limits_{(i,j)\in w}[(\delta_{p_{i\rightarrow j}^{0,1}}^{1}+\delta_{p_{i\rightarrow j}^{0,1}}^{0.5})*(\delta_{p_{j\leftarrow i}^{0,1}}^{1}+\delta_{p_{j\leftarrow i}^{0,1}}^{0.5})-\delta_{p_{i\rightarrow j}^{0,1}}^{0.5}\delta_{p_{j\leftarrow i}^{0,1}}^{0.5}]
\end{split}
\end{equation}
The warning propagation convergence speed is very fast, and gives the same results as BP, but it does not converge when the mean variable degree is greater than 3.704 on the ER random graph. For some single graphs, the convergence degree is greater than 3.704, but it is very close to 3.704 in most single graph networks.
\section{One Step Replica Symmetry Breaking Theory}
In this section, we introduce 1RSB theory for spin glass systems, which is calculated using the graph expansion method. We first introduce the generalized partition function, free energy, SP, grand free energy and complexity in the general case. To obtain the ground state energy,
we must consider the limiting behavior of the DMDS problem at $\beta=\infty$, so we next derive the simplified equations at $\beta=\infty$ for the DMDS problem,
and then introduce the numerical simulation process for population dynamics.
\subsection{General One step Replica Symmetry Breaking Theory}
The RS theory only finds low energy configurations. To study the subspace structure for a given problem, researchers have been developing 1RSB theory. In 1RSB theory, our order parameter is a free energy function. At higher temperatures, the microscopic thermodynamic state consisting of some higher energy configurations determines the statistical physics properties of the given system, and the subspace of this microscopic state is ergodic. However, at lower temperatures the microscopic state is no longer ergodic but is divided into several subspaces, and the contribution of these
subspaces to the equilibrium properties are not the same. 
We define the generalized partition function $\Xi$ by
\begin{equation}
\Xi(y;\beta)=\sum_{\alpha}e^{-yF_{0}^{\alpha}(\beta)},
\end{equation}
We use $\alpha$ to denote the microscope states that achieve the minimum free energy, and thus $F_{0}^{\alpha}$ has the following form:

\begin{equation}
F_{0}^{(\alpha)}=\sum_{i}f_{i} - \sum_{(i,j)}f_{(i,j)}.
\end{equation}
Four cavity messages $p_{i\rightarrow j}\equiv (p_{i\rightarrow j}^{0,0},p_{i\rightarrow j}^{0,1},p_{i\rightarrow j}^{1,0},p_{i\rightarrow j}^{1,1},)$ are defined on the edges $(i,j)$ in a given graph, and the cavity messages $p_{i\rightarrow j}$ averaged on solution clusters are denoted by $P_{i\rightarrow j}(p_{i\rightarrow j})$. This has the iteration equation

\begin{equation}
\begin{split}
P_{i\rightarrow j}(p)&=\frac{\prod_{k\in\partial i^{+}}\int\mathcal{D}p_{k\rightarrow i}P_{k\rightarrow i}(p)e^{-yf_{i\rightarrow j}}}{\prod_{k\in\partial i^{+}}\int\mathcal{D}p_{k\rightarrow i}P_{k\rightarrow i}(p)e^{-yf_{i\rightarrow j}}}\\
&\times\frac{\prod_{k\in\partial i^{-}\backslash j}\int\mathcal{D}p_{k\leftarrow i}P_{k\leftarrow i}(p)\delta(p_{i\rightarrow j}-I_{i\rightarrow j}[p_{\partial i\backslash j}])}{\prod_{k\in\partial i^{-}\backslash j}\int\mathcal{D}p_{k\leftarrow i}P_{k\leftarrow i}(p)}
\end{split}
\end{equation}

\begin{equation}
\begin{split}
P_{i\leftarrow j}(p)&=\frac{\prod_{k\in\partial i^{-}}\int\mathcal{D}p_{k\leftarrow i}P_{k\leftarrow i}(p)e^{-yf_{i\leftarrow j}}}{\prod_{k\in\partial i^{-}}\int\mathcal{D}p_{k\leftarrow i}P_{k\leftarrow i}(p)e^{-yf_{i\leftarrow j}}}\\
&\times\frac{\prod_{k\in\partial i^{+}\backslash j}\int\mathcal{D}p_{k\rightarrow i}P_{k\rightarrow i}(p)\delta(p_{i\leftarrow j}-I_{i\leftarrow j}[p_{\partial i\backslash j}])}{\prod_{k\in\partial i^{+}\backslash j}\int\mathcal{D}p_{k\rightarrow i}P_{k\rightarrow i}(p)}
\end{split}
\end{equation}

The notation $I_{i\rightarrow j}[p_{\partial i\backslash j}]$ is short-hand for messages updating equation (2),
and $I_{i\leftarrow j}[p_{\partial i\backslash j}]$ is short-hand for messages updating equation (3).
The weight free energies $f_{i\rightarrow j}$ and $f_{i\leftarrow j}$ are respectively equal to
\begin{equation}
f_{i\rightarrow j}=-\frac{1}{\beta}\ln\{\sum\limits_{c_{i}}e^{-\beta c_{i}}[\prod\limits_{k\in\partial i^{+}}\sum\limits_{c_{k}}p_{k\rightarrow i}^{c_{k}, c_{i}}-\delta_{c_{i}}^{0}\prod\limits_{k\in\partial i^{+}}p_{k\rightarrow i}^{0,0}]\prod\limits_{l\in\partial i^{-}\backslash j}\sum\limits_{c_{l}}p_{l\leftarrow i}^{c_{l}, c_{i}}\}
\end{equation}
\begin{equation}
f_{i\leftarrow j}=-\frac{1}{\beta}\ln\{\sum\limits_{c_{i}}e^{-\beta c_{i}}[\prod\limits_{k\in\partial i^{+}\backslash j}\sum\limits_{c_{j}}p_{k\rightarrow i}^{c_{k}, c_{i}}-\delta_{c_{i}+c_{j}}^{0}\prod\limits_{k\in\partial i^{+}\backslash j}p_{k\rightarrow i}^{0,0}]\prod\limits_{l\in\partial i^{-}}\sum\limits_{c_{l}}p_{l\leftarrow i}^{c_{l}, c_{i}}\}
\end{equation}
The generalized free energy density $g_{0}$ is
\begin{equation}
g_{0}\equiv \frac{G_{0}}{N}=\frac{\sum_{i}g_{i}-\sum_{(i,j)}g_{(i,j)}}{N}
\end{equation}
where
\begin{equation}
g_{i}=\frac{1}{y}\ln[\prod_{j\in\partial i}\int\mathcal{D}p_{i\rightarrow j}P_{i\rightarrow j}(p)e^{-yf_{i}}]
\end{equation}
\begin{sequation}
g_{(i,j)}=\frac{1}{y}\ln[\int\int\mathcal{D}p_{i\rightarrow j}\mathcal{D}p_{j\rightarrow i}P_{i\rightarrow j}(p)P_{j\leftarrow i}(p)e^{-yf_{(i,j)}}]
\end{sequation}

We further have the mean free energy density

\begin{equation}
<f>\equiv \frac{F}{N}=\frac{\sum_{i}<f_{i}>-\sum_{(i,j)}<f_{(i,j)}>}{N}
\end{equation}

where

\begin{equation}
<f_{i}>=\frac{\prod_{j\in\partial i^{+}}\int\mathcal{D}p_{i\leftarrow j}P_{i\leftarrow j}(p)\prod_{j\in\partial i^{-}}\int\mathcal{D}p_{i\rightarrow j}P_{i\rightarrow j}(p)e^{-yf_{i}}f_{i}}{\prod_{j\in\partial i^{+}}\int\mathcal{D}p_{i\leftarrow j}P_{i\leftarrow j}(p)\prod_{j\in\partial i^{-}}\int\mathcal{D}p_{i\rightarrow j}P_{i\rightarrow j}(p)e^{-yf_{i}}}
\end{equation}
\begin{equation}
<f_{(i,j)}>=\frac{\int\int\mathcal{D}p_{i\rightarrow j}\mathcal{D}p_{j\rightarrow i}P_{i\rightarrow j}(p)P_{j\rightarrow i}(p)e^{-yf_{(i,j)}}f_{(i,j)}}{\int\int\mathcal{D}p_{i\rightarrow j}\mathcal{D}p_{j\rightarrow i}P_{i\rightarrow j}(p)P_{j\rightarrow i}(p)e^{-yf_{(i,j)}}}
\end{equation}

Finally, with mean free energy $<f>$ and generalized free energy $g$, we derive the complexity as
$\sum(y)=y(<f>-g)$.

\subsection{Coarse-Grain Survey Propagation }
We next derive the SP for the case $\beta=\infty$ and estimate the ground state energy, and then predict the energy density using the SPD method. We find that the SP results fit with the SPD results, which are as good as the belief propagation decimation (BPD) results.
To obtain the SP, we must to know the form of the free energy $F_{i\rightarrow j}$ at a
temperature of zero. From the general form, we can derive the free energy $F_{i\rightarrow j}$ as

\begin{equation}
\begin{split}
F_{i\leftarrow j}&=-\frac{1}{\beta}\ln\{[2\prod_{k\in\partial i^{+}\backslash j}(p_{k\rightarrow i}^{0,0}+p_{k\rightarrow i}^{1,0})-\prod_{k\in\partial i^{+}\backslash j}p_{k\rightarrow i}^{0,0}]\prod_{k\in\partial i^{-}}(p_{k\leftarrow i}^{0,0}+p_{k\leftarrow i}^{1,0})\\
&+2e^{-\beta}\prod_{k\in\partial i^{+}\backslash j}(p_{k\rightarrow i}^{0,1}+p_{k\rightarrow i}^{1,1})\prod_{k\in\partial i^{-}}(p_{k\leftarrow i}^{0,1}+p_{k\leftarrow i}^{1,1})\}
\end{split}
\end{equation}
\begin{equation}
\begin{split}
F_{i\rightarrow j}&=-\frac{1}{\beta}\ln\{2[\prod_{k\in\partial i^{+}}(p_{k\rightarrow i}^{0,0}+p_{k\rightarrow i}^{1,0})-\prod_{k\in\partial i^{+}}p_{k\rightarrow i}^{0,0}]\prod_{k\in\partial i^{-}\backslash j}(p_{k\leftarrow i}^{0,0}+p_{k\leftarrow i}^{1,0})\\
&+2e^{-\beta}\prod_{k\in\partial i^{+}}(p_{k\rightarrow i}^{0,1}+p_{k\rightarrow i}^{1,1})\prod_{k\in\partial i^{-}\backslash j}(p_{k\leftarrow i}^{0,1}+p_{k\leftarrow i}^{1,1})\}
\end{split}
\end{equation}
\begin{equation}
F_{i\leftarrow j}=\Theta[(\sum\limits_{k\in\partial i^{+}\backslash j}\delta_{p_{k\rightarrow i}^{0,1}}^{1}+\sum\limits_{k\in\partial i^{-}}\delta_{p_{k\leftarrow i}^{0,1}}^{1})-1]
\end{equation}
\begin{equation}
F_{i\rightarrow j}=\Theta[(\sum\limits_{k\in\partial i^{+}}\delta_{p_{k\rightarrow i}^{0,1}}^{1}+\sum\limits_{k\in\partial i^{-}\backslash j}\delta_{p_{k\leftarrow i}^{0,1}}^{1})-1]+\delta(\sum\limits_{k\in\partial i^{+}}\delta_{p_{k\rightarrow i}^{0,1}}^{0.5},k^{+})
\end{equation}
the survey propagation for general case as
\begin{equation}
\begin{split}
P_{i\rightarrow j}(p)&=\frac{\prod_{k\in\partial i^{+}}\int\mathcal{D}p_{k\rightarrow i}P_{k\rightarrow i}(p)e^{-yf_{i\rightarrow j}}}{\prod_{k\in\partial i^{+}}\int\mathcal{D}p_{k\rightarrow i}P_{k\rightarrow i}(p)e^{-yf_{i\rightarrow j}}}\\
&\times\frac{\prod_{k\in\partial i^{-}\backslash j}\int\mathcal{D}p_{k\leftarrow i}P_{k\leftarrow i}(p)\delta(p_{i\rightarrow j}-I_{i\rightarrow j}[p_{\partial i\backslash j}])}{\prod_{k\in\partial i^{-}\backslash j}\int\mathcal{D}p_{k\leftarrow i}P_{k\leftarrow i}(p)}
\end{split}
\end{equation}

\begin{equation}
\begin{split}
P_{i\leftarrow j}(p)&=\frac{\prod_{k\in\partial i^{-}}\int\mathcal{D}p_{k\leftarrow i}P_{k\leftarrow i}(p)e^{-yf_{i\leftarrow j}}}{\prod_{k\in\partial i^{-}}\int\mathcal{D}p_{k\leftarrow i}P_{k\leftarrow i}(p)e^{-yf_{i\leftarrow j}}}\\
&\times\frac{\prod_{k\in\partial i^{+}\backslash j}\int\mathcal{D}p_{k\rightarrow i}P_{k\rightarrow i}(p)\delta(p_{i\leftarrow j}-I_{i\leftarrow j}[p_{\partial i\backslash j}])}{\prod_{k\in\partial i^{+}\backslash j}\int\mathcal{D}p_{k\rightarrow i}P_{k\rightarrow i}(p)}
\end{split}
\end{equation}

We can obtain the SP at a temperature of zero using equations (28)-(31) as

\begin{equation}
P_{i\leftarrow j}(\delta_{p_{i\leftarrow j}^{0,1}}^{1})=\frac{\prod\limits\limits_{k\in\partial i^{+}\backslash j}P_{k\rightarrow i}(\delta_{p_{k\rightarrow i}^{0,1}}^{0.5})\prod\limits\limits_{k\in\partial i^{-}}[1-P_{k\leftarrow i}(\delta_{p_{k\leftarrow i}^{0,1}}^{1})]}{\prod\limits\limits_{k\in\partial i^{+}\backslash j}[1-P_{k\rightarrow i}(\delta_{p_{k\rightarrow i}^{0,1}}^{1})]\prod\limits\limits_{k\in\partial i^{-}}[1-P_{k\leftarrow i}(\delta_{p_{k\leftarrow i}^{0,1}}^{1})]+e^{-y}P_{01}^{1}}
\end{equation}
where$P_{01}^{1}=1-\prod\limits\limits_{k\in\partial i^{+}\backslash j}[1-P_{k\rightarrow i}(\delta_{p_{k\rightarrow i}^{0,1}}^{1})]\prod\limits\limits_{k\in\partial i^{-}}[1-P_{k\leftarrow i}(\delta_{p_{k\leftarrow i}^{0,1}}^{1})]$
\begin{equation}
P_{i\leftarrow j}(\delta_{p_{i\leftarrow j}^{0,1}}^{0.5})=\frac{\{\prod\limits\limits_{k\in\partial i^{+}\backslash j}[1-P_{k\rightarrow i}(\delta_{p_{k\rightarrow i}^{0,1}}^{1})]-\prod\limits\limits_{k\in\partial i^{+}\backslash j}P_{k\rightarrow i}(\delta_{p_{k\rightarrow i}^{0,1}}^{0.5})\}\prod\limits\limits_{k\in\partial i^{-}}[1-P_{k\leftarrow i}(\delta_{p_{k\leftarrow i}^{0,1}}^{1})]}{\prod\limits\limits_{k\in\partial i^{+}\backslash j}[1-P_{k\rightarrow i}(\delta_{p_{k\rightarrow i}^{0,1}}^{1})]\prod\limits\limits_{k\in\partial i^{-}}[1-P_{k\leftarrow i}(\delta_{p_{k\leftarrow i}^{0,1}}^{1})]+e^{-y}P_{01}^{1}}
\end{equation}
\begin{equation}
P_{i\rightarrow j}(\delta_{p_{i\rightarrow j}^{0,1}}^{0.5})=\frac{\{\prod\limits\limits_{k\in\partial i^{+}}[1-P_{k\rightarrow i}(\delta_{p_{k\rightarrow i}^{0,1}}^{1})]-\prod\limits\limits_{k\in\partial i^{+}}P_{k\rightarrow i}(\delta_{p_{k\rightarrow i}^{0,1}}^{0.5})\}\prod\limits\limits_{k\in\partial i^{-}\backslash j}[1-P_{k\leftarrow i}(\delta_{p_{k\leftarrow i}^{0,1}}^{1})]}{\{\prod\limits\limits_{k\in\partial i^{+}}[1-P_{k\rightarrow i}(\delta_{p_{k\rightarrow i}^{0,1}}^{1})]-\prod\limits\limits_{k\in\partial i^{+}}P_{k\rightarrow i}(\delta_{p_{k\rightarrow i}^{0,1}}^{0.5})\}\prod\limits\limits_{k\in\partial i^{-}\backslash j}[1-P_{k\leftarrow i}(\delta_{p_{k\leftarrow i}^{0,1}}^{1})]+e^{-y}P_{01}^{-1}}
\end{equation}

where$P_{01}^{-1}=1-\{\prod\limits\limits_{k\in\partial i^{+}}[1-P_{k\rightarrow i}(\delta_{p_{k\rightarrow i}^{0,1}}^{1})]-\prod\limits\limits_{k\in\partial i^{+}}P_{k\rightarrow i}(\delta_{p_{k\rightarrow i}^{0,1}}^{0.5})\}\prod\limits\limits_{k\in\partial i^{-}\backslash j}[1-P_{k\leftarrow i}(\delta_{p_{k\leftarrow i}^{0,1}}^{1})]$
These two equations are the SP at a temperature of zero. In the same way, we can derive the free energy of node $i$ and edge $(i,j)$ as

\begin{equation}
f_{i}=\Theta[(\sum\limits_{j\in\partial i^{+}}\delta_{p_{j\rightarrow i}^{0,1}}^{1}+\sum\limits_{j\in\partial i^{-}}\delta_{p_{j\leftarrow i}^{0,1}}^{1})-1]+\delta(\sum\limits_{j\in\partial i^{+}}\delta_{p_{j\rightarrow i}^{0,1}}^{0.5},k^{+})
\end{equation}
\begin{equation}
f_{(i,j)}=\sum\limits_{(i,j)\in w}[(\delta_{p_{i\rightarrow j}^{0,1}}^{1}+\delta_{p_{i\rightarrow j}^{0,1}}^{0.5})*(\delta_{p_{j\leftarrow i}^{0,1}}^{1}+\delta_{p_{j\leftarrow i}^{0,1}}^{0.5})-\delta_{p_{i\rightarrow j}^{0,1}}^{0.5}\delta_{p_{j\leftarrow i}^{0,1}}^{0.5}]
\end{equation}

and the grand free energy of node $i$ and edge $(i,j)$ for the general case as

\begin{equation}
g_{i}=\frac{1}{y}\ln[\prod_{k\in\partial i^{+}}\int\mathcal{D}p_{k\rightarrow i}P_{k\rightarrow i}(p)e^{-yf_{i}}
\prod_{k\in\partial i^{-}}\int\mathcal{D}p_{k\leftarrow i}P_{k\leftarrow i}(p)]
\end{equation}
\begin{equation}
g_{(i,j)}=\frac{1}{y}\ln[\int\int\mathcal{D}p_{i\rightarrow j}\mathcal{D}p_{j\leftarrow i}P_{i\rightarrow j}(p)P_{j\leftarrow i}(p)e^{-yf_{(i,j)}}]
\end{equation}

From equations (35)-(38), we can derive the grand free energy of node $i$ and edge $(i,j)$ at a temperature of zero as

\begin{equation}
\begin{split}
G_{i}&=-\frac{1}{y}\sum\limits_{i=1}^{N}\ln\{(1-e^{-y})\{\prod\limits\limits_{k\in\partial i^{+}}[1-P_{k\rightarrow i}(\delta_{p_{k\rightarrow i}^{0,1}}^{1})]-\prod\limits\limits_{k\in\partial i^{+}}P_{k\rightarrow i}(\delta_{p_{k\rightarrow i}^{0,1}}^{0.5})\}\\
&\times\prod\limits\limits_{k\in\partial i^{-}}[1-P_{k\leftarrow i}(\delta_{p_{k\leftarrow i}^{0,1}}^{1})]+e^{-y}\}
\end{split}
\end{equation}
\begin{equation}
G_{(i,j)}=-\frac{1}{y}\sum\limits_{i=1}^{N}\ln\{1-(1-e^{-y})p_{i\rightarrow j}^{0,1}[nh]\}
\end{equation}
where
\begin{equation}
\begin{split}
p_{i\rightarrow j}^{0,1}[nh]=&[P_{i\rightarrow j}(\delta_{p_{i\rightarrow j}^{0,1}}^{0.5})+P_{i\rightarrow j}(\delta_{p_{i\rightarrow j}^{0,1}}^{1})]\\
&\times[P_{j\leftarrow i}(\delta_{p_{j\leftarrow i}^{0,1}}^{0.5})+P_{j\leftarrow i}(\delta_{p_{j\leftarrow i}^{0,1}}^{1})]\\
&-P_{i\rightarrow j}(\delta_{p_{i\rightarrow j}^{0,1}}^{0.5}) P_{j\leftarrow i}(\delta_{p_{j\leftarrow i}^{0,1}}^{0.5})
\end{split}
\end{equation}
the grand free energy density is
\begin{equation}
g_{0}\equiv \frac{G_{0}}{N}=\frac{\sum_{i}g_{i}-\sum_{(i,j)}g_{(i,j)}}{N}
\end{equation}

The free energy of the macroscopic state $\alpha$ when $\beta=\infty$ has several different ground state energies $E_{min}$, we cannot calculate the ground state energies one at a time. We are only concerned with the average ground state energy, so we focus our discussion on this. We denote the microscopic average minimal energy by $<E_{\beta=\infty}>$, which is calculated using the following equation:

\begin{equation}
\begin{split}
&<E_{\beta=\infty}>=\frac{\partial (yG_{0})}{\partial y}\\
&=\sum\limits_{i=1}^{N}\frac{e^{-y}-e^{-y}\{\prod\limits\limits_{k\in\partial i^{+}}[1-P_{k\rightarrow i}(\delta_{p_{k\rightarrow i}^{0,1}}^{1})]-\prod\limits\limits_{k\in\partial i^{+}}P_{k\rightarrow i}(\delta_{p_{k\rightarrow i}^{0,1}}^{0.5})\}
\prod\limits\limits_{k\in\partial i^{-}}[1-P_{k\leftarrow i}(\delta_{p_{k\leftarrow i}^{0,1}}^{1})]}{(1-e^{-y})\{\prod\limits\limits_{k\in\partial i^{+}}[1-P_{k\rightarrow i}(\delta_{p_{k\rightarrow i}^{0,1}}^{1})]-\prod\limits\limits_{k\in\partial i^{+}}P_{k\rightarrow i}(\delta_{p_{k\rightarrow i}^{0,1}}^{0.5})\}
\prod\limits\limits_{k\in\partial i^{-}}[1-P_{k\leftarrow i}(\delta_{p_{k\leftarrow i}^{0,1}}^{1})]+e^{-y}}\\
&-\sum\limits_{(i,j)\in\partial w}^{N}\frac{e^{-y}p_{i\rightarrow j}^{0,1}[nh]}{1-(1-e^{-y})p_{i\rightarrow j}^{0,1}[nh]}
\end{split}
\end{equation}

We can study the ensemble average properties of the DMDS problem using population dynamics and equations (28)-(31), (37) and (38). Graph 1 indicates the results of the ensemble everage 1RSB population dynamics for the DMDS problem on the ER random graph with mean connectivity C=5. The complexity $\sum=0$ at $y=0$, and the complexity is not a monotonic function of the Parisi parameter $y$. It increases as the Parisi parameter $y$ increases, and reaches its maximum value when $y\approx 3.7$. The complexity then begins to decrease as $y$ increases and becomes negative when $y\approx 8.35$. From Graph 1, we can see that there are two parts to the complexity graph when it is a function of energy, but because of only the concave part is decline function of energy, so it has the physical meaning. The grand free energy is not a monotonic function of $y$ either. It reach a maximum when the complexity becomes negative at $y\approx 8.35$, so the corresponding energy density $u=0.3212$ is the minimum energy density for the DMDS problem at this mean connectivity.\\

\begin{figure}[htb]
  \centering
  \includegraphics[width=12cm,height=7cm]{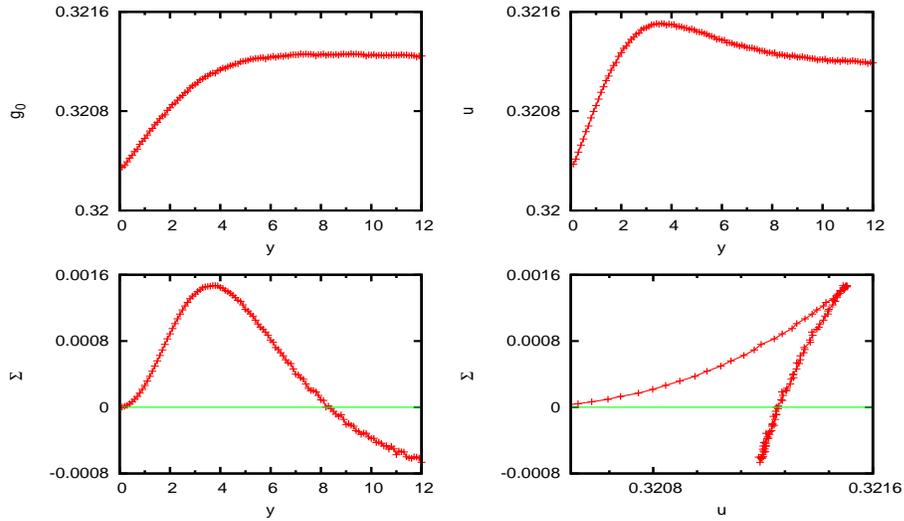}
  \caption{The 1RSB results for the zero temperature DMDS problem on the ER random graph with mean connectivity $c=5$ using population dynamics. In the upper two and bottom left graphs, the $x$-axis denotes the Parisi parameter $Y$, and the $y$-axis denotes the thermodynamic quantities.The complexity becomes negative when the Parisi parameter is approximately $8.35$. At this point, we select the corresponding energy as the ground state energy, which equals 0.3212. In the bottom right graph, the $x$-axis denotes the energy density and the $y$-axis denotes the complexity.}
\end{figure}

We can calculate some microscopic statistical quantities using equations (32-34) at a temperature of zero. For example, for the probability (statistical total weight of all macrostates) of the variable staying in a coarse grained state, we use $p_{i}(0)$ to denote the probability of the variable staying in a totally uncovered state, $p_{i}(1)$ to denote the probability of the variable staying in a totally covered state, and $p_{i}(*)$ to denote the probability of the variable staying in an unfrozen (some microstates covered) state. We can derive the representation of these three probabilities using 1RSB mean field theory as

\begin{equation}
P_{i}(0)=\frac{\{\prod\limits\limits_{k\in\partial i^{+}}[1-P_{k\rightarrow i}(\delta_{p_{k\rightarrow i}^{0,1}}^{1})]-\prod\limits\limits_{k\in\partial i^{+}}P_{k\rightarrow i}(\delta_{p_{k\rightarrow i}^{0,1}}^{0.5})\}\prod\limits\limits_{k\in\partial i^{-}}[1-P_{k\leftarrow i}(\delta_{p_{k\leftarrow i}^{0,1}}^{1})]}{\{\prod\limits\limits_{k\in\partial i^{+}}[1-P_{k\rightarrow i}(\delta_{p_{k\rightarrow i}^{0,1}}^{1})]-\prod\limits\limits_{k\in\partial i^{+}}P_{k\rightarrow i}(\delta_{p_{k\rightarrow i}^{0,1}}^{0.5})\}\prod\limits\limits_{k\in\partial i^{-}}[1-P_{k\leftarrow i}(\delta_{p_{k\leftarrow i}^{0,1}}^{1})]+e^{-y}P_{01}^{-1}}
\end{equation}
where$P_{01}^{-1}=1-\{\prod\limits\limits_{k\in\partial i^{+}}[1-P_{k\rightarrow i}(\delta_{p_{k\rightarrow i}^{0,1}}^{1})]-\prod\limits\limits_{k\in\partial i^{+}}P_{k\rightarrow i}(\delta_{p_{k\rightarrow i}^{0,1}}^{0.5})\}\prod\limits\limits_{k\in\partial i^{-}}[1-P_{k\leftarrow i}(\delta_{p_{k\leftarrow i}^{0,1}}^{1})]$
\begin{equation}
P_{i}(*)=\frac{\sum\limits_{j\in\partial i^{+}}P_{j\rightarrow i}(\delta_{p_{j\rightarrow i}^{0,1}}^{1})plp+\sum\limits_{j\in\partial i^{-}}P_{j\leftarrow i}(\delta_{p_{j\rightarrow i}^{0,1}}^{1})pln+\prod\limits\limits_{k\in\partial i^{+}}P_{k\rightarrow i}(\delta_{p_{k\rightarrow i}^{0,1}}^{0.5})\prod\limits\limits_{k\in\partial i^{-}}[1-P_{k\leftarrow i}(\delta_{p_{k\leftarrow i}^{0,1}}^{1})]}{\{\prod\limits\limits_{k\in\partial i^{+}}[1-P_{k\rightarrow i}(\delta_{p_{k\rightarrow i}^{0,1}}^{1})]-\prod\limits\limits_{k\in\partial i^{+}}P_{k\rightarrow i}(\delta_{p_{k\rightarrow i}^{0,1}}^{0.5})\}\prod\limits\limits_{k\in\partial i^{-}}[1-P_{k\leftarrow i}(\delta_{p_{k\leftarrow i}^{0,1}}^{1})]+e^{-y}P_{01}^{-1}}
\end{equation}
where $plp=\prod\limits\limits_{k\in\partial i^{+}\backslash j}[1-P_{k\rightarrow i}(\delta_{p_{k\rightarrow i}^{0,1}}^{1})]\prod\limits\limits_{k\in\partial i^{-}}[1-P_{k\leftarrow i}(\delta_{p_{k\leftarrow i}^{0,1}}^{1})]$ and the
$pln=\prod\limits\limits_{k\in\partial i^{+}}[1-P_{k\rightarrow i}(\delta_{p_{k\rightarrow i}^{0,1}}^{1})]\prod\limits\limits_{k\in\partial i^{-}\backslash j}[1-P_{k\leftarrow i}(\delta_{p_{k\leftarrow i}^{0,1}}^{1})]$
\begin{equation}
P_{i}(1)=1-P_{i}(0)-P_{i}(*)
\end{equation}
Using the 1RSB population dynamics, we obtain the minimal energy densities for ER random network ensembles with different mean connectivities .In Table 1, we list the theoretical computational results for $C\le 10$. We
can see that the ground state energy and transition point depend on the mean connectivity $C$. From \cite{22}, we know that the transition point does not depend on the mean connectivity $C$ in undirected networks.
\begin{table}[!hbp]
\tiny
\caption{The  parisi parameter $y^{*}$ transition point and the ground state energy for ER random graphs.}
\begin{tabular}{p{1.3cm}p{0.9cm}p{0.9cm}p{0.9cm}p{0.9cm}p{0.9cm}p{0.9cm}p{0.8cm}p{0.8cm}}
\hline
C & 3 & 4 & 5 & 6&7&8&9&10 \\
\hline
$y^{*}\approx$ & $\infty$ & $> 30$ & 8.35 & 7.9&7.65&7.45&7.38&7.35 \\
\hline
$u_{min}(1RSB)$ & 0.4421 & 0.3726& 0.3212 & 0.2828&0.2532&0.2298&0.2109&0.1953 \\
\hline
$u_{min}(RS)$ & 0.4358 & 0.3676 & 0.3182 & 0.2808&0.2519&0.2291&0.2102&0.1947 \\
\hline
\end{tabular}
\end{table}
\\
In these simulations, we update the population $M_{I}=1000$ times. That is, we update each element in the population $1000$ times on average to reach a stable point for the population, and sample $M_{S}=5000$ times to obtain the transition point of $\sum$ (second $\sum=0$ points) and the corresponding ground state energy value $E_{min}$ on the ER random graph. The cluster transition point is only correct when the Parisi parameter sample distance is $\bigtriangledown y\ge 0.1$, but the ground state energy is correct in any small enough Parisi parameter sampling distance with a precision of $\bigtriangledown E=0.0001$. We use two types of population, a positive population and a negative population, and set the population size to $N=1000000$. Increasing the number of updates or the number of samples does not affect the simulation results. However, increasing the population size $N$ used to calculate the thermodynamic quantities improves the results. Within a sampling distance of $\bigtriangledown y=0.1$, we can also obtain good results with fewer updates. However, with a sampling distance of $\bigtriangledown y=0.01$, we need an increasing number of updates to obtain good results. The required numbers of updates and samples increases as the variable degree decreases.
\subsection{Survey Propagation Decimation}
We studied undirected networks using SP in \cite{22}. We can also study the statistical properties of microscopic configurations in a single directed network using the SP in equations (32)-(34). The results are similar to those for an undirected network. For example, SP can find a stable point for a given directed network easily when the Parisi parameter $y$ is small enough, and we can then calculate the thermodynamic quantities using equations (35), (36), and (39)-(42). However, SP does not converge when $y$ is too large. For example, SP does not converge when $y\ge 7$ for an ER random network when $C=10$. We have already discussed the reasons for this in undirected network in \cite{22}, namely that the coarse-grained assumption and 1RSB mean field theory are not sufficient to describe microscopic configuration spaces when the energy is close to the ground state energy, so a more detailed coarse-grained assumption and a higher-level expansion of the partition function are required. For further discussion of convergence of coarse-grained SP, see \cite{26,27}.\\
We can construct one or more solutions that are close to the optimal DMDS for a given graph $W$ using 1RSB mean field theory. One very efficient algorithm is the SPD algorithm \cite{22,26,27}. The main idea of this algorithm is to first determine the probability of being covered, and then select a small subset of variables that has the highest probability of being covered. We then set the covering probability for this subset of variables to $ c_{i}=1$, and delete all variables for which the covering probability equals 1, along with the adjacent edges, and then simplify the network iteratively. If a node $i$ is unobserved (it is empty and has no adjacent occupied parent node), then the output messages $P_{i\rightarrow j}$ and $P_{i\leftarrow j}$ are updated according to equations (32)-(34), On the other hand, if node $i$ is empty but observed (it has at least one adjacent occupied parent node), then this node presents no restriction on the occupation states of its unoccupied parent neighbors. For such a node $i$, the output messages $P_{i\rightarrow j}$ and $P_{i\leftarrow j}$ are then updated according to the following equations:
\begin{equation}
P_{i\leftarrow j}(\delta_{p_{i\leftarrow j}^{0,1}}^{1})=0;
\end{equation}
\begin{equation}
P_{i\leftarrow j}(\delta_{p_{i\leftarrow j}^{0,1}}^{0.5})=\frac{\{\prod\limits\limits_{k\in\partial i^{+}\backslash j}[1-P_{k\rightarrow i}(\delta_{p_{k\rightarrow i}^{0,1}}^{1})]\}\prod\limits\limits_{k\in\partial i^{-}}[1-P_{k\leftarrow i}(\delta_{p_{k\leftarrow i}^{0,1}}^{1})]}{\prod\limits\limits_{k\in\partial i^{+}\backslash j}[1-P_{k\rightarrow i}(\delta_{p_{k\rightarrow i}^{0,1}}^{1})]\prod\limits\limits_{k\in\partial i^{-}}[1-P_{k\leftarrow i}(\delta_{p_{k\leftarrow i}^{0,1}}^{1})]+e^{-y}P_{01}^{1}}
\end{equation}
where $P_{01}^{1}=1-\prod\limits\limits_{k\in\partial i^{+}\backslash j}[1-P_{k\rightarrow i}(\delta_{p_{k\rightarrow i}^{0,1}}^{1})]\prod\limits\limits_{k\in\partial i^{-}}[1-P_{k\leftarrow i}(\delta_{p_{k\leftarrow i}^{0,1}}^{1})]$,
\begin{equation}
P_{i\rightarrow j}(\delta_{p_{i\rightarrow j}^{0,1}}^{0.5})=\frac{\{\prod\limits\limits_{k\in\partial i^{+}}[1-P_{k\rightarrow i}(\delta_{p_{k\rightarrow i}^{0,1}}^{1})]\}\prod\limits\limits_{k\in\partial i^{-}\backslash j}[1-P_{k\leftarrow i}(\delta_{p_{k\leftarrow i}^{0,1}}^{1})]}{\{\prod\limits\limits_{k\in\partial i^{+}}[1-P_{k\rightarrow i}(\delta_{p_{k\rightarrow i}^{0,1}}^{1})]\}\prod\limits\limits_{k\in\partial i^{-}\backslash j}[1-P_{k\leftarrow i}(\delta_{p_{k\leftarrow i}^{0,1}}^{1})]+e^{-y}P_{01}^{-1}}
\end{equation}
where $P_{01}^{-1}=1-\{\prod\limits\limits_{k\in\partial i^{+}}[1-P_{k\rightarrow i}(\delta_{p_{k\rightarrow i}^{0,1}}^{1})]\}\prod\limits\limits_{k\in\partial i^{-}\backslash j}[1-P_{k\leftarrow i}(\delta_{p_{k\leftarrow i}^{0,1}}^{1})]$.\\
As with equations (47)-(49), the marginal probability distribution $P_{i}$ for an observed empty node $i$ can be evaluated according to
\begin{equation}
P_{i}(0)=\frac{\{\prod\limits\limits_{k\in\partial i^{+}}[1-P_{k\rightarrow i}(\delta_{p_{k\rightarrow i}^{0,1}}^{1})]\}\prod\limits\limits_{k\in\partial i^{-}}[1-P_{k\leftarrow i}(\delta_{p_{k\leftarrow i}^{0,1}}^{1})]}{\{\prod\limits\limits_{k\in\partial i^{+}}[1-P_{k\rightarrow i}(\delta_{p_{k\rightarrow i}^{0,1}}^{1})]\}\prod\limits\limits_{k\in\partial i^{-}}[1-P_{k\leftarrow i}(\delta_{p_{k\leftarrow i}^{0,1}}^{1})]+e^{-y}P_{01}^{-1}}
\end{equation}
where$P_{01}^{-1}=1-\{\prod\limits\limits_{k\in\partial i^{+}}[1-P_{k\rightarrow i}(\delta_{p_{k\rightarrow i}^{0,1}}^{1})]\}\prod\limits\limits_{k\in\partial i^{-}}[1-P_{k\leftarrow i}(\delta_{p_{k\leftarrow i}^{0,1}}^{1})]$
\begin{equation}
P_{i}(*)=\frac{\sum\limits_{j\in\partial i^{+}}P_{j\rightarrow i}(\delta_{p_{j\rightarrow i}^{0,1}}^{1})plp+\sum\limits_{j\in\partial i^{-}}P_{j\leftarrow i}(\delta_{p_{j\rightarrow i}^{0,1}}^{1})pln}{\{\prod\limits\limits_{k\in\partial i^{+}}[1-P_{k\rightarrow i}(\delta_{p_{k\rightarrow i}^{0,1}}^{1})]\}\prod\limits\limits_{k\in\partial i^{-}}[1-P_{k\leftarrow i}(\delta_{p_{k\leftarrow i}^{0,1}}^{1})]+e^{-y}P_{01}^{-1}}
\end{equation}
where $plp=\prod\limits\limits_{k\in\partial i^{+}\backslash j}[1-P_{k\rightarrow i}(\delta_{p_{k\rightarrow i}^{0,1}}^{1})]\prod\limits\limits_{k\in\partial i^{-}}[1-P_{k\leftarrow i}(\delta_{p_{k\leftarrow i}^{0,1}}^{1})]$ and the
$pln=\prod\limits\limits_{k\in\partial i^{+}}[1-P_{k\rightarrow i}(\delta_{p_{k\rightarrow i}^{0,1}}^{1})]\prod\limits\limits_{k\in\partial i^{-}\backslash j}[1-P_{k\leftarrow i}(\delta_{p_{k\leftarrow i}^{0,1}}^{1})]$
\begin{equation}
P_{i}(1)=1-P_{i}(0)-P_{i}(*)
\end{equation}
We now present the details of this algorithm.\\

(1) Read the network $W$, set the covering probability of every vertex to uncertain, and define four coarse-grained messages, $P_{i\rightarrow j}(\delta_{p_{i\rightarrow j}^{0,1}}^{0.5}),P_{i\rightarrow j}(\delta_{p_{i\rightarrow j}^{0,1}}^{1}),\\
 P_{j\rightarrow i}(\delta_{p_{j\rightarrow i}^{0,1}}^{0.5})$, and $P_{j\rightarrow i}(\delta_{p_{j\rightarrow i}^{0,1}}^{1})$, on each edge of the given graph. Randomly initialize the messages in the interval (0,1], ensuring that for every pair of messages $P_{i\rightarrow j}(\delta_{p_{i\rightarrow j}^{0,1}}^{0.5})$,
$P_{i\rightarrow j}(\delta_{p_{i\rightarrow j}^{0,1}}^{1})$ and $P_{j\rightarrow i}(\delta_{p_{j\rightarrow i}^{0,1}}^{0.5})$, $P_{j\rightarrow i}(\delta_{p_{j\rightarrow i}^{0,1}}^{1})$, the normalization conditions $P_{i\rightarrow j}(\delta_{p_{i\rightarrow j}^{0,1}}^{1})$+$P_{i\rightarrow j}(\delta_{p_{i\rightarrow j}^{0,1}}^{0.5})$
$+P_{i\rightarrow j}(\delta_{p_{i\rightarrow j}^{0,1}}^{0.25})$+$P_{i\rightarrow j}(\delta_{p_{i\rightarrow j}^{0,1}}^{0})=1$ and $P_{i\leftarrow j}(\delta_{p_{i\leftarrow j}^{0,1}}^{1})$+$P_{i\leftarrow j}(\delta_{p_{i\leftarrow j}^{0,1}}^{0.5})$+$P_{i\leftarrow j}(\delta_{p_{i\leftarrow j}^{0,1}}^{0.25})$+$P_{i\leftarrow j}(\delta_{p_{i\leftarrow j}^{0,1}}^{0})=1$ are satisfied. An appropriate setting for the macroscopic inverse temperature $y$ is a value close to the threshold value. For example, if SP does not converge when $y\ge 3.01$, then we set $y=3$.\\
(2) Iterate the coarse-grained SP equations (equations (32)-(34) or equations (47)-(49)), for $L_{0}$ steps, aiming for convergence to a stable point. At each iteration, select one node $i$ and update all messages corresponding to node $i$. After updating the messages $L_{0}$ times, calculate the coarse-grained probability $(P_{i}(1),P_{i}(*),P_{i}(0))$ using equations (44)-(46) or equations (50)-(52).\\
(3) Sort the variables that are not frozen in descending order according to the value of $P_{i}(1)$. Select the first $r$ percent to set the covering state as $c_{i}=1$, and add these variables to the DMDS.\\
(4) Simplify the network by deleting all the edges between the observed nodes and deleting all the occupied variables. If the remaining network still contains one or more leaf nodes \cite{20}, then apply the GLR process \cite{21} until there are no leaf nodes in the network and simplify the network again. This procedure is repeated (simplify-GLR-simplify) until the network contains no leaf nodes. If the network contains no nodes and no edges, then stop the program and output the DMDS.\\
(5) If the network still contains some nodes, then iterate equations (32)-(34) or equations (47)-(49) for $L_{1}$ steps. Repeat steps (3), (4), and (5).\\

Figure 2 shows the numerical results of the SPD algorithm on an ER random graph. We can see that the SPD results are very close to the BPD results, and thus the SPD algorithm finds an almost optimal solution. We perform the BPD algorithm as detailed in\cite{20}.

\begin{figure}[htbp]
  \centering
  \includegraphics[width=8cm,height=5cm]{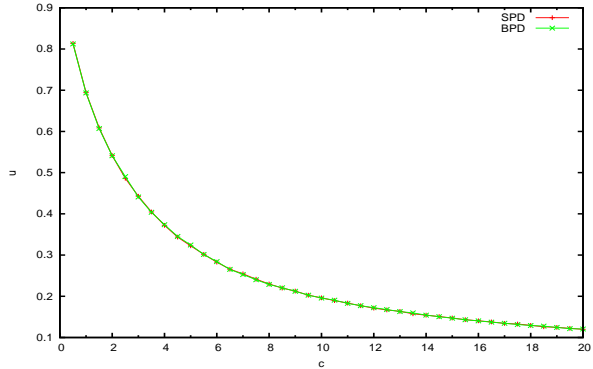}
  \caption{The solid line is the result of the SPD algorithm, and the crosses are the results of the BPD algorithm. Our simulation was performed on an ER random graph with $10^4$ variables.}
\end{figure}
\section{$\hspace{1.5mm}$Discussion}
In this work, we first derived the warning propagation and proved that the warning propagation equation only converges when the network does not contain a core \cite{21}.There is only one warning in the vertex cover problem \cite{28},but the MDS problem has two warnings. In the DMDS problem, a positive edge $p_{i\leftarrow j}$ has two warnings, but a negative edge $p_{i\rightarrow j}$ has one warning. The reason for this is that each nodes only requires the parent nodes, not child nodes, to be occupied to be able to observe itself, so only the psitive edges have first type warnings. Second, we derived the SP equation at a temperature of zero to find the ground state energy and the corresponding transition point of the macroscopic inverse temperature. The change rules of the transition point does not like with undirected MDS problem. It is a monotonic function of the mean variable degree in the DMDS problem, but not in the undirected MDS problem \cite{22}. The corresponding energy of the transition point Parisi parameter $Y$ equals the threshold value $x_{c}$, namely $E_{Y}=x_{c}$. We then implemented the SPD algorithm at a temperature of zero to estimate the size of a DMDS. The results are as good as the BPD results.\\
We have previously studied the MDS problem on undirected networks and directed networks using statistical physics, and more recently we studied the undirected MDS problem using 1RSB mean field theory. We have now studied the DMDS problem using 1RSB theory. We plan to study the MDS and DMDS problems using long range frustration theory in the future.

\section{$\hspace{2mm}$ Acknowledgement}
Yusupjan Habibulla thanks Prof. Haijun Zhou for helpful discussions and guidance. Yusupjan Habibulla also thanks Prof. Xiaosong Chen for helpful discussions and support. We implemented the numerical simulations on the cluster in the School of Physics and Technology at Xinjiang University. This research was supported by the doctoral startup fund of Xinjiang University of China (grant number 208-61357) and the national natural science foundation of china (grant number 11765021). We thank Peter Humphries, PhD, from Edanz Group (www.edanzediting.com/ac) for editing a draft of this manuscript.

\end{document}